\pdfoutput=1
\RequirePackage{ifpdf}
\ifpdf 
\documentclass[pdftex]{sigma}
\else
\documentclass{sigma}
\fi

\numberwithin{equation}{section}

\begin{document}

\newcommand{\arXivNumber}{1504.03705}

\allowdisplaybreaks

\renewcommand{\thefootnote}{$\star$}

\renewcommand{\PaperNumber}{057}

\FirstPageHeading

\ShortArticleName{Racah Polynomials and Recoupling Schemes of $\mathfrak{su}(1,1)$}

\ArticleName{Racah Polynomials\\ and Recoupling Schemes of $\boldsymbol{\mathfrak{su}(1,1)}$\footnote{This paper is a~contribution to the Special Issue on Exact Solvability and Symmetry Avatars
in honour of Luc Vinet.
The full collection is available at
\href{http://www.emis.de/journals/SIGMA/ESSA2014.html}{http://www.emis.de/journals/SIGMA/ESSA2014.html}}}

\Author{Sarah POST}
\AuthorNameForHeading{S.~Post}
\Address{Department of Mathematics, University of Hawai`i at M\=anoa, Honolulu, HI,  96822, USA}
\Email{\href{ mailto:spost@hawaii.edu}{spost@hawaii.edu}}

\ArticleDates{Received April 16, 2015, in f\/inal form July 14, 2015; Published online July 23, 2015}

\Abstract{The connection between the recoupling scheme of four copies of $\mathfrak{su}(1,1)$, the generic superintegrable system on the 3 sphere, and bivariate Racah polynomials is identif\/ied. The Racah polynomials are presented as connection coef\/f\/icients between eigenfunctions separated in dif\/ferent spherical coordinate systems and equivalently as dif\/ferent irreducible decompositions of the tensor product representations. As a consequence of the model, an extension of the quadratic algebra ${\rm QR}(3)$ is given. It is shown that this algebra closes only with the inclusion of an additional shift operator, beyond the eigenvalue operators for the bivariate Racah polynomials, whose polynomial eigenfunctions are determined. The duality between the variables and the degrees, and hence the bispectrality of the polynomials, is interpreted in terms of expansion coef\/f\/icients of the separated solutions.}

\Classification{33C45; 33D45; 33D80; 81R05; 81R12}

\renewcommand{\thefootnote}{\arabic{footnote}}
\setcounter{footnote}{0}

\section{Introduction}
The connection between group theory, special functions and orthogonal polynomials is an area that has been of signif\/icant interest for many years now and has yielded many beautiful results fundamental in theory and in application. In this work, we give a Lie algebraic description of Tratnik's extension of the Racah polynomials \cite{Trat1991}. This description relies on the connection between these polynomials and representations of the quadratic algebra associated with the `generic' superintegrable system on the three sphere~\cite{KMPost11} and its realization in terms of  positive discrete series representations of $\mathfrak{su}(1,1)$ obtained by Genest and Vinet~\cite{genest2014generic}. The system is given by the following superintegrable quantum Hamiltonian
\begin{gather} \label{H}
H=-\Delta_{S_3} +\frac{a_1}{s_1^2}+\frac{a_2}{s_2^2}+\frac{a_3}{s_3^2}+\frac{a_4}{s^2_4}, \qquad s_1^2+s_2^2+s_3^2+s_4^2=1,
\end{gather}
and is referred to a generic since all non-degenerate,  second-order superintegrable systems on conformally f\/lat pseudo-Euclidean spaces of three dimensions can be obtained through appropriate limits from this system~\cite{CapelKressPost2015}. It admits 6 linearly independent constants of the motion and hence is considered superintegrable~\cite{superreview}.

Remarkably, Genest and Vinet~\cite{genest2014generic} have shown that this system can be realized as the tensor product of four copies of  positive discrete series representations of $\mathfrak{su}(1,1)$. They have used this representation to identify the $9j$ symbols of the algebra and derive identities for the symbols using the separated solutions of the Hamiltonian~(\ref{H}) in cylindrical coordinates, which are given in terms of Jacobi polynomials. The $9j$ symbols are given by rational functions multiplied by the vacuum coef\/f\/icients.

In this paper, we discuss the same tensor product representation but instead focus on bases which separate in spherical coordinates and their associated coupling schemes. We show that bivariate Racah polynomials can be obtained as expansion coef\/f\/icients between two spherical coordinate systems, or equivalently, two dif\/ferent coupling schemes. The bispectrality of the polynomials is obtained naturally from the model. We show also that there is another set of commuting dif\/ference operators for the bivariate Racah polynomials which f\/its naturally into the scheme. When this dif\/ference operator is added to the algebra, it closed to form a quadratic algebra on 9 generators, which we will call~${\rm QR}(9)$ as it is an extension of the Racah algebra~${\rm QR}(3)$ from the univariate case.

The remainder of the paper is organized as follows. Section~\ref{section2} contains the necessary background material on the two-sphere case and the realization of univariate Racah polynomials as expansion coef\/f\/icients between dif\/ferent bases for three copies of $\mathfrak{su}(1,1)$. Section~\ref{section3} contains the main results of the paper, namely the connection between the bivariate Racah polynomials and the coupling schemes of four copies of $\mathfrak{su}(1,1)$. Section~\ref{section4} contains the results concerning the additional dif\/ference operator and the algebra ${\rm QR}(9)$ as well as an interpretation of the bispectrality properties of the Racah polynomials in terms of the tensor product representation.  Section~\ref{section5} brief\/ly gives some conclusions and future outlook.

\section{Background}\label{section2}
\subsection[Positive discrete series representations of  $\mathfrak{su}(1,1)$]{Positive discrete series representations of  $\boldsymbol{\mathfrak{su}(1,1)}$}\label{section2.1}

Consider the following operator realization of the positive discrete series representation of $\mathfrak{su}(1,1)$
\begin{gather}
K_0=\frac14 \left(-\partial_s ^2 +s^2+\frac{a}{s^2}\right),\qquad K_{\pm}=\frac14 \left( (s\mp \partial_s)^2-\frac{a}{s^2}\right), \label{discreterep}
\end{gather}
satisfying the $\mathfrak{su}(1,1)$ commutation relations
\begin{gather}
\label{su11}
 [K_0, K_{\pm}]=\pm K_{\pm}, \qquad [K_-, K_+]=2K_0
 \end{gather}
with Casimir operator
\begin{gather*}
Q=K_0^2-\frac12 \{K_+, K_-\}=\nu(\nu-1), \qquad a=\frac14(4\nu-1)(4\nu-3).
\end{gather*}
Here $\{ A, B\}=AB+BA.$
As is well known, see, e.g., \cite{goodman2009symmetry,humphreys1972introduction}, the tensor product of multiple copies of such representations will itself be a positive discrete series representation of~$\mathfrak{su}(1,1)$. For example, the tensor product of two such representations would have generators
\begin{gather*}
K_\mu^{(12)} =K_\mu^{(1)}+K_{\mu}^{(2)},  \qquad \mu=0,+,-,
\end{gather*}
where the upper indices indicate the factor in the tensor product.  A subscript is also  appended to the variable to distinguish each factor in the representation, i.e.,
\begin{gather*}
K_0^{(i)}=\frac12 \left(-\partial_{s_i} ^2 +s_i^2+\frac{a_i}{s_i^2}\right),
\end{gather*}
and similarly for $K_{\pm}^{(i)}$.

The tensor product can be decomposed into irreducible components. In the case of the tensor product of two representations, the irreducible components are indexed by a non-negative integer~$x$ such that the total Casimir
\begin{gather*}
Q^{(12)}=\big(K_0^{(12)}\big)^2-\frac12 \big\{K_+^{(12)}, K_-^{(12)}\big\}
\end{gather*}
takes the  values $\nu_{12}(\nu_{12}-1)$ with $\nu_{12}=\nu_{1}+\nu_2+x$, on each irreducible component. In the case of two or more tensor products, the decomposition is not canonical. For example for three components,
$V=V^{(1)}\otimes V^{(2)} \otimes V^{(3)},$
with algebra generators
\begin{gather*}
K_\mu^{(123)} =K_\mu^{(1)}+K_{\mu}^{(2)}+K_{\mu}^{(3)},  \qquad \mu=0,+,-,
\end{gather*}
it is possible to decompose f\/irst by coupling $V^{(1)}$ and $V^{(2)}$ into components of~$V^{(12)}$  with each irreducible component indexed by the parameter~$\nu_{12}$ and then adding the third component. The basis in this coupling scheme would be $|\nu_{12}, \nu\rangle  $ with action of operators
\begin{alignat}{3}
&  Q^{(12)} |\nu_{12}, \nu\rangle  =\nu_{12}(\nu_{12}-1) |\nu_{12}, \nu\rangle , \qquad && \nu_{12}=\nu_1+\nu_2+x, & \nonumber\\
       \label{Qval}  & Q  |\nu_{12}, \nu\rangle  =\nu(\nu-1) |\nu_{12}, \nu\rangle , \qquad && \nu=\nu_1+\nu_2+\nu_3+N.&
\end{alignat}
 Here $Q$ is the  Casimir operator of the triple tensor  product representation of $\mathfrak{su}(1,1)$, called the total Casimir operator,  given by
 \begin{gather}\label{Ctot}
 Q=\big(K_0^{(123)}\big)^2-\frac12 \big\{K_+^{(123)}, K_-^{(123)}\big\}.
 \end{gather}
This operator takes the values as in (\ref{Qval}), indexed by non-negative integer $N$, on irreducible components.
Coupling in the other direction, by reducing f\/irst $V^{(2)}\otimes V^{(3)}$ into irreducible components indexed by $\nu_{23}$, depending on non-negative integer $n$, and then into components indexed by $N$ determined by the  total Casimir operator,  gives the following alternate basis
\begin{alignat*}{3}
&  Q^{(23)} |\nu_{23}, \nu\rangle  =\nu_{23}(\nu_{23}-1) |\nu_{23}, \nu\rangle , \qquad && \nu_{23}=\nu_2+\nu_3+n, &\nonumber\\
& Q  |\nu_{23}, \nu\rangle  =\nu(\nu-1) |\nu_{23}, \nu\rangle , \qquad && \nu=\nu_1+\nu_2+\nu_3+N.&
\end{alignat*}
Note that from its def\/inition~(\ref{Ctot}),  the total Casimir operator can be expressed as a linear combination of the intermediate Casimirs
\begin{gather}\label{totalQ}
Q=Q^{(123)}=Q^{(12)}+Q^{(13)}+Q^{(23)}-Q^{(1)}-Q^{(2)}-Q^{(3)}, \qquad Q^{(i)}=\nu_i(\nu_i-1).
\end{gather}
The extension to 4 copies of $\mathfrak{su}(1,1)$ representations is clear. In this case, the total Casimir operator has the form
\begin{gather*}
Q\equiv Q^{(1234)}=\sum Q^{(ij)}-2\sum Q^{(i)}.
\end{gather*}

Using the operator realization of the positive discrete series representation of $\mathfrak{su}(1,1)$~(\ref{discreterep}), we can identify the total Casimir operator $Q$ with the Hamiltonian in (\ref{H}) and the intermediate Casimirs with the integrals of the motion for the Hamiltonian. Indeed the Hamiltonian~$H$ admits 6~linearly independent integrals of the motion coinciding exactly with the intermediate Casimir operators in the discrete series representation~(\ref{discreterep}), namely
\begin{gather} \label{Qijmodel}
Q^{(ij)}=-\frac14 \left(s_i\partial_{s_j}-s_j\partial_{s_i}\right)^2+\frac{a_i s_j^2}{4s_i^2}+\frac{a_j s_i^2}{4s_j^2}+\frac{a_i+a_j-1}{4}. \end{gather}
The two-sphere version of~(\ref{H}) is
\begin{gather*} 
H^{(123)}=-\Delta_{S_2} +\frac{a_1}{s_1^2}+\frac{a_2}{s_2^2}+\frac{a_3}{s_3^2}, \qquad s_1^2+s_2^2+s_3^2=1,
\end{gather*}
and can be expressed as
\begin{gather*}
H^{(123)}=4Q^{(123)}+\frac34
\end{gather*}
and hence satisf\/ies $[H^{(123)}, Q^{(ij)}]=0.$ For the three-sphere model, arising from the tensor product of four $\mathfrak{su}(1,1)$ representations, the Hamiltonian~(\ref{H}) is related to the total Casimir via
\begin{gather*}
H=4Q^{(1234)}, \qquad \big[H, Q^{(ij)}\big]=0.
\end{gather*}

\subsection[Three products of $\mathfrak{su}(1,1)$ and the Racah algebra]{Three products of $\boldsymbol{\mathfrak{su}(1,1)}$ and the Racah algebra}

The decomposition of three products of $\mathfrak{su}(1,1)$ and its connection with the Racah algebra was discovered by Genest, Vinet and Zhedanov \cite{genest2013superintegrability}. The algebra and its connection with Racah polynomials will be fundamental for the following sections, so we recall the results here.

Consider the following interbasis expansion coef\/f\/icients
\begin{gather*}
P_{\nu_{23}, \nu_{12}}= \langle \nu_{23}, \nu| \nu_{12}, \nu\rangle,
\end{gather*}
and def\/ine the following operators:
\begin{gather} \label{k1}
k_1 P_{\nu_{23}, \nu_{12}}=-\frac{1}2\big\langle \nu_{23}, \nu| Q^{(12)} |\nu_{12}, \nu\big\rangle
\\ \label{k2}
k_2 P_{\nu_{23}, \nu_{12}}=-\frac{1}2\big\langle \nu_{23}, \nu| Q^{(23)} |\nu_{12},\nu\big\rangle.
\end{gather}
Here, as in the remainder of the paper, the operators~$Q^{(ij)}$ are interpreted as acting on the vector space~$V$ while the operators~$k_j$, and later $K_j,$ as acting on the expansion coef\/f\/icients.

{\sloppy To determine the operators $k_1$, we see that  $Q^{(12)}$ acts on the right, i.e., on the basis $|\nu_{12}, \nu_{123}, \nu\rangle$ giving
\begin{gather*}
k_1 P_{\nu_{23}, \nu_{12}}=-\frac{1}{2}\nu_{12}(\nu_{12}-1) P_{\nu_{12}, \nu_{23}},
\end{gather*}
but  $Q^{(12)}$ will also act on the left, i.e., on the  $|\nu_{23}, \nu\rangle$ basis,  as
\begin{gather} \label{recur}
k_1 P_{\nu_{23}, \nu_{12}}= \sum_{\mu} C_{\mu} \langle \mu, \nu| \nu_{12}, \nu\rangle,
\end{gather}
for some expansion coef\/f\/icients $C_{\mu}$.
Similarly, the operator $k_2$ will satisfy
\begin{gather*} k_2 P_{\nu_{23}, \nu_{12}}=-\frac{1}{2} \nu_{23}(\nu_{23}-1) P_{\nu_{12}, \nu_{23}}, \qquad
k_2 P_{\nu_{23}, \nu_{12}}= \sum_{\mu} D_{\mu} \langle \nu_{23}, \nu| \mu, \nu\rangle.
\end{gather*}
In order to determine these expansion coef\/f\/icients and to hence use the recurrence rela\-tion~(\ref{recur})
we will identify the algebra generated by $k_1$ and $k_2$ with the quadratic Racah algebra ${\rm QR}(3)$~\mbox{\cite{granovskii1992mutual, Zhedanov1991}}.
Indeed, computing the commutator of~$k_1$ and ~$k_2$ leads to a third operator $k_3$ def\/ined as
\begin{gather*}
k_3=[k_1, k_2].
\end{gather*}
The equivalent operator in terms of the intermediate Casimir operators is def\/ined as follows
\begin{gather*}
R\equiv \big[Q^{(12)},  Q^{(23)}\big],
\end{gather*}
so that
\begin{gather*}
k_3 P_{\nu_{23}, \nu_{12}} = \frac14\big\langle \nu_{23}, \nu| \big[Q^{(12)}, Q^{(23)}\big]|\nu_{12}, \nu\big\rangle
 = \frac14\langle \nu_{23}, \nu| R |\nu_{12}, \nu\rangle .
 \end{gather*}
By direct computation using the $\mathfrak{su}(1,1)$ algebra relations~(\ref{su11}) and the def\/inition of the intermediate Casimir operators (\ref{discreterep}), we see that $R$ can be expressed as the sum over permutations of~$\{1,2,3\}$ of cubic terms in the~$\mathfrak{su}(1,1)$ generators,
\begin{gather} \label{Rdef}
R=\sum2 \epsilon_{ijk} K_0^{(i)}K_-^{(j)}K_+^{(k)}.
\end{gather}
Here $\epsilon_{ijk} $ is the sign of the permutation $(ijk)$. This formula holds in general; namely for the operators~$R_{ijk}$ def\/ined via
\begin{gather} \label{Rjs}
R_{ijk}=\epsilon_{ijk}\big[Q^{(ij)},Q^{(jk)}\big],
\end{gather}
the identity~(\ref{Rdef}) holds with $R$ replaced by~$R_{ijk}$. In the notation of this section $R\equiv R_{123}$.

}

To complete the algebra relations, it remains only to compute the commutators $[k_1, k_3]$ and $[k_3, k_2]$, which are  determined using  the algebra relations for the $Q^{(ij)}$
\begin{gather*}
\big[R, Q^{(12)}\big]=-2\big(Q^{(12)}\big)^2-2\big\{Q^{(12)},Q^{(23)}\big\}\\
\hphantom{\big[R, Q^{(12)}\big]=}{} +2\big(Q+Q^{(1)}+Q^{(2)}+Q^{(3)}\big)Q^{(12)}+2\big(Q^{(1)}-Q^{(2)}\big)\big(Q^{(3)}-Q\big)
\\
\big[Q^{(23)}, R\big]
=-2\big(Q^{(23)}\big)^2-2\big\{Q^{(12)},Q^{(23)}\big\}\\
\hphantom{\big[Q^{(23)}, R\big]=}{}
+2\big(Q+Q^{(1)}+Q^{(2)}+Q^{(3)}\big)Q^{(23)}+2\big(Q^{(2)}-Q^{(3)}\big)\big(Q^{(1)}-Q\big).
\end{gather*}
These relations can be obtained either through the $\mathfrak{su}(1,1)$ algebra relations or, more practically, using the models (\ref{discreterep}).
In terms of the $k_j$ operators,  the quadratic algebra relations are
\begin{gather}
[k_1, k_2]=k_3,\qquad
    [ k_2, k_3 ]=k_2^2+\{k_1, k_2\}+dk_2+e_1, \nonumber\\
[k_3, k_1]=k_1^2+\{k_1, k_2\}+dk_1+e_2,\label{QR3}
\end{gather}
with
\begin{gather*} d=\frac12\big(Q^{(1)}+Q^{(2)}+Q^{(3)}+Q^{(123)}\big), \\
e_{1}=-\frac{1}4\big(Q^{(3)}-Q^{(2)})(Q^{(1)}-Q^{(123)}\big), \qquad
e_{2}=\frac{1}4\big(Q^{(2)}-Q^{(1)}\big)\big(Q^{(3)}-Q^{(123)}\big).
\end{gather*}
From these algebra relations~(\ref{QR3}) it is possible to determine the action of~$k_1$ as a shift operator in the~$\nu_{23}$, or equivalently the~$n$ index, as well as the action of~$k_2$ as a shift operators in the~$\nu_{12}$, or equivalently the~$x$ index. Indeed, this direct approach was used in~\cite{KMPost11}. However, guided by hindsight, we see that this algebra is precisely the algebra~${\rm QR}(3)$ associated with the Racah polynomials and we has make use of this identif\/ication to determine the action of the opera\-tors~$k_1$ and~$k_2$ in terms of the three-term recurrence relation and eigenvalue equations for the Racah polynomials.

Consider now the Racah polynomials def\/ined by
\begin{gather*}
r_n(\alpha, \beta, \gamma, \delta;x)=(\alpha+1)_n(\beta+\delta+1)_n(\gamma+1)_n\\
\hphantom{r_n(\alpha, \beta, \gamma, \delta,x)=}{}\times {}_4F_3\left[\begin{matrix} -n, n+\alpha+\beta+1, -x, x+\gamma +\delta+1\\ \alpha+1, \beta+\delta+1, \gamma+1\end{matrix}; 1\right],
\end{gather*}
which are polynomials of  degree $n$ in $\lambda(x)=x(x+\gamma+\delta+1)$.  For consistency with the following sections, we shall parametrize   the polynomials  as
\begin{gather*} r_n(\beta_1-\beta_0-1, \beta_2-\beta_1-1, -N-1,N+\beta_1; x)\\
\qquad =(\beta_1-\beta_0)_n(-N)_n(N+\beta_2)_n  \, {}_4F_3\left[\begin{matrix}-n, n+\beta_2-\beta_0-1, -x, x+\beta_1\\ \beta_1-\beta_0, N+\beta_2, -N\end{matrix}; 1\right].
\end{gather*}
The polynomials  satisfy the following eigenvalue equation{\samepage
\begin{gather}
\Lambda(x, \beta, N)r_n(x) \equiv  \left[-\mathcal{L}(x, \beta; N) +\left(\frac{\beta_2-\beta_0}{2}\right)\left(\frac{\beta_2-\beta_0}{2}-1\right)\right]r_n(x)\nonumber\\
\hphantom{\Lambda(x, \beta, N)r_n(x)}{} =\kappa\left(n, \frac{\beta_2-\beta_0}{2}\right)r_n(x), \label{Lambda}
\end{gather}}

\noindent
with
\begin{gather*}
\mathcal{L}(x, \beta, N)=\big[ B(x)(T_x-1)+E(x)\big(T_x^{-1}-1\big)
\big]R_n(x),\\
 B(x)=\frac{(x+\beta_1-\beta_0)(x+\beta_1)(x+\beta_2+N)(N-x)}{(2x+\beta_1)(2x+\beta_1+1)},\\  
 E(x)=\frac{x(x+\beta_0)(N-x-\beta_1+\beta_2)(N+x+\beta_1)}{(2x+\beta_1)(2x+\beta_1-1)}, \qquad 
 \kappa(n, c)=  (n+c ) (n+c-1 ).
 \end{gather*}
The operator $\mathcal{L}(x, \beta, N)$ is of the form  given in~\cite{GI2010}. Note that the function $\kappa$ is related to the values of the Casimir operators as follows
\begin{gather*}
\nu_{12}(\nu_{12}-1) =\kappa(x, \nu_1+\nu_2), \qquad \nu_{23}(\nu_{23}-1)=\kappa(n, \nu_2+\nu_3).
\end{gather*}

The operators associated with multiplication by the transform variable and the eigenvalue equation for the Racah polynomials,
\begin{gather} \label{k12R}
k_1^{\rm R}=-\frac{1}{2}\kappa\left(x, \frac{\beta_1+1}{2}\right), \qquad k_2^{\rm R}=-\frac{1}{2}\Lambda(x, \beta, N),
\end{gather}
gives a realization of the algebra ${\rm QR}(3)$~(\ref{QR3}) with the following parameters
\begin{gather*} 
d=\frac12\left(N(N+\beta_2)+\frac12\beta_0(\beta_0-\beta_1+1)
+\frac12 \beta_1(\beta_1-\beta_2)+\frac12\beta_2(\beta_2-1)-\frac12\right),\\
e_1=-\frac{1}{4}\left(\frac{\beta_2-\beta_0}{2}-1\right)\left(\beta_1+\beta_0-\frac{\beta_2-\beta_0}{2}\right)
\left(N+\frac{\beta_2-\beta_0}{2}\right)\left(N+\beta_0+\frac{\beta_2-\beta_0}{2}\right),\\
e_2=\frac{1}{4}\left(\frac{\beta_1+1}{2}-1\right) \left(\beta_0+1-\frac{\beta_1+1}{2}\right)  \left(N+\frac{\beta_1+1}{2}\right)\left(N+\beta_2-\frac{\beta_1+1}{2}\right).
\end{gather*}
Finally, we are in a position to determine the  expansion coef\/f\/icients $P_{\nu_{23}, \nu_{12}}$ as Racah polynomials and identify the operators $k_1$ and $k_2$ (\ref{k1}),~(\ref{k2}) with $k_1^{\rm R}$ and~$k_2^{\rm R}$~(\ref{k12R}). The identif\/ication is accomplished by taking
\begin{gather}
\beta_0=2\nu_1-1, \qquad \beta_1=2\nu_1+2\nu_2-1, \qquad \beta_2=2\nu_1+2\nu_2+2\nu_3-1. \label{coeffsRx}
\end{gather}
 The identif\/ication is only determined up to a possible choice of conjugation, namely
\begin{gather*}
k_i^{\rm R}=G(x) k_i G(x)^{-1}.
\end{gather*}
To determine the necessary gauge, it is important to use the requirement that the basis be normalized. Therefore, we chose
\begin{gather*}
P_{\nu_{23}, \nu_{12}}= G(x,n)R_n(x),\end{gather*}
with
\begin{gather*} G(x,n)^2  =  P_{0, \nu_{12}} P_{\nu_{23}, 0}\\
\hphantom{G(x,n)^2}{}
=\frac{(-N)_x(2\nu_2)_x(2\nu_1+2\nu_2-1)_x(N+2\nu_1+2\nu_2+2\nu_3-1)_x(\nu_1+\nu_2+\frac12)_x}{(2\nu_1)_x(-N-2\nu_3+1)_x(\nu_1+\nu_2-\frac12)_x (N+2\nu_1+2\nu_2)_x x!}\\
\hphantom{G(x,n)^2  =}{}
\times \frac{n!(2\nu_3)_n (n+2\nu_2+2\nu_3-1)_n(N+2\nu_2+2\nu_3)_n(-N-2\nu_1+1)_n}{(-N)_n(2\nu_2+2\nu_3)_{2n}(2\nu_2)_n(N+2\nu_1+2\nu_2+2\nu_3-1)_n }\\
\hphantom{G(x,n)^2  =}{}
\times \frac{(2\nu_2+2\nu_3)_N (-N-2\nu_1-2\nu_2+1)_N}{(-N-2\nu_1+1)_N(2\nu_3)_N}.
\end{gather*}

The action of the operator $Q^{(23)}$ can then be determined as follows
\begin{gather*}
k_2 P_{\nu_{23}, \nu_{12}}=k_2 G(x,n)R_n(x)=G(x,n) k_2^{\rm R} R_n(x),
\end{gather*}
and so
\begin{gather*}
k_2= G(x,n) k_2^{\rm R} G(x,n)^{-1}.
\end{gather*}
Thus f\/inally, we see the action of $Q^{(23)}$ on the basis $|\nu_{12}, \nu\rangle$ is given by
\begin{gather} \label{Q23final}
Q^{(23)}= G(x,n) \Lambda(x, \beta, N)G(x,n)^{-1}\equiv \hat{\Lambda}(x, \beta, N),
\end{gather}
where $T_x|\nu_{12}, \nu\rangle=|\nu_{12}+1, \nu\rangle$ and similarly for $T_x^{-1},$ where the notation makes use of the fact that $\nu_{12}$ is linear in $x$.

\section[Four products of $\mathfrak{su}(1,1)$ and bivariate Racah polynomials]{Four products of $\boldsymbol{\mathfrak{su}(1,1)}$ and bivariate Racah polynomials}\label{section3}

In this section we consider four products of $\mathfrak{su}(1,1)$ and determine their relation to the bivariate polynomials obtained by Tratnik~\cite{Trat1991}. As above, we obtain the polynomials through interbasis expansion coef\/f\/icients and identify them by comparing the algebra of the operators with the algebra generated by the recurrence operators for the polynomials.

We take as our initial bases the vectors obtained by coupling f\/irst $V^{(1)}$ with~$V^{(2)}$ to obtain components~$V^{(12)}$ and then coupling this with $V^{(3)}$ to obtain components $V^{(123)}$ and then f\/inally with~$V^{(4)}$ to obtain a complete basis indexed by $\nu_{12}$, $\nu_{123}$ and $\nu$, depending on non-negative integers~$x_1$, $x_2 $ and~$N$,  satisfying
\begin{alignat}{3}
& Q^{(12)}|\nu_{12}, \nu_{123}, \nu\rangle =\nu_{12}(\nu_{12}-1)  |\nu_{12}, \nu_{123}, \nu\rangle, \qquad && \nu_{12}=\nu_1+\nu_2+x_1,&\nonumber\\
&Q^{(123)}|\nu_{12}, \nu_{123}, \nu\rangle =\nu_{123}(\nu_{123}-1)  |\nu_{12}, \nu_{123}, \nu\rangle, \qquad && \nu_{123}=\nu_1+\nu_2+\nu_3+x_2,& \label{12123} \\
&Q|\nu_{12}, \nu_{123}, \nu\rangle =\nu(\nu-1)  |\nu_{12}, \nu_{123}, \nu\rangle, \qquad && \nu=\nu_1+\nu_2+\nu_3+\nu_4+N.\nonumber
\end{alignat}
We will build up the action of each of the intermediate Casimir operators~$Q^{(ij)}$ on this basis using the quadratic algebra generated by these 6 operators plus their four linearly independent commutator, the~$R_{ijk}$'s~(\ref{Rjs}), for example
\begin{gather*}
R_{123}=\big[Q^{(12)},Q^{(23)}\big]=\big[Q^{(13)},Q^{(12)}\big]=\big[Q^{(23)}, Q^{(13)}\big].
\end{gather*}

The remaining algebra relations are obtained through direct, though tedious computations using the def\/initions of~$R_{ijk}$ and~$Q^{(ij)}$. They are as follows
 \begin{gather}
 \big[R_{jk\ell}, Q^{(ij)}\big] = \big\{Q^{(ik)}, Q^{(j\ell)}\big\}-\big\{Q^{(i\ell)}, Q^{(jk)}\big\}-2\big(Q^{(i)}-Q^{(j)}\big)\big(Q^{(k)}-Q^{(\ell)}\big)\nonumber\\
\hphantom{\big[R_{jk\ell}, Q^{(ij)}\big] =}{} +2\big(Q^{(j)}+Q^{(k)}\big)Q^{(i\ell)}+2\big(Q^{(i)}+Q^{(\ell)}\big)Q^{(jk)}
\nonumber\\
\hphantom{\big[R_{jk\ell}, Q^{(ij)}\big] =}{}-2\big(Q^{(j)}+Q^{(\ell)}\big)Q^{(ik)}-2\big(Q^{(i)}+Q^{(k)}\big)Q^{(j\ell)},\label{quadsameijQijk}
\\
\label{quadsameij} \big[R_{ijk}, Q^{(ij)}\big] =  \big\{ Q^{(ij)}, Q^{(ik)}\big\}-\big\{Q^{(ij)}, Q^{(jk)}\big\} -2\big(Q^{(i)}-Q^{(j)}\big)\big(Q^{(ijk)}-Q^{(k)}\big).
\end{gather}
Notice that for a f\/ixed $ijk$, the quadratic algebra relations~(\ref{quadsameij}) can be expressed via $Q^{(ijk)}$ as
\begin{gather*}
\big[R_{ijk}, Q^{(ij)}\big] = -2\big(Q^{(ij)}\big)^2-2\big\{Q^{(ij)}, Q^{(jk)}\big\} +2\big(Q^{(ijk)}+Q^{(i)}+Q^{(j)}+Q^{(k)}\big)Q^{(ij)} \\
\hphantom{\big[R_{ijk}, Q^{(ij)}\big] =}{} -2\big(Q^{(i)}-Q^{(j)}\big)\big(Q^{(ijk)}-Q^{(k)}\big)
\end{gather*}
and hence replicate the Racah algebra relations on components with f\/ixed values of the Casi\-mir~$Q^{(ijk)}$ by taking $k_1=-Q^{(ij)}/2 $ and $ k_2=-Q^{(jk)}/2.$  This subalgebra will be utilized to obtain the action of the operator~$Q^{(23)}$ on the basis~(\ref{12123}).

\subsection[The basis for $\big\{ Q^{(12)}, Q^{(23)}\big\}$ and univariate Racah polynomials]{The basis for $\boldsymbol{\big\{ Q^{(12)}, Q^{(23)}\big\}}$ and univariate Racah polynomials}

Let us now consider the basis for $V$ composed of eigenvectors for the commuting operators $Q^{(23)}$ and $Q^{(123)}$, namely $|\nu_{23}, \nu_{123}, \nu\rangle$ as well as the action of $Q^{(23)}$ on the basis~$|\nu_{12}, \nu_{123}, \nu\rangle. $ In order to determine that action of $Q^{(23)}$ on this basis, we note that~$Q^{(23)}$ also commutes with~$Q^{(123)}$ and so the action of $Q^{(23)}$ will be just to shift the index $\nu_{12}$. Indeed the action will be the same as the product of three representations and so the operator $Q^{(23)}$ will be exactly the eigenvalue shift operator of the Racah polynomials given in~(\ref{Q23final}) except that the parameter~$\nu_{123}$ is no longer the total Casimir and is instead given by $\nu_{123}=\nu_1+\nu_2+\nu_3+x_2.$ Thus, the action of~$Q^{(23)}$ on the basis is given by~(\ref{Q23final}) with parameters chosen as in~(\ref{coeffsRx}) with the integer $N$ is replaced by~$x_2$.
The expansion coef\/f\/icients for these two bases are then
\begin{gather*}
\langle \nu_{23}, \nu_{123}, \nu|\nu_{12}, \nu_{123}, \nu\rangle
 =\sqrt{\omega_{x_1}\sigma_{n_1}} r_{n_1}(2\nu_2-1, 2\nu_3-1, -x_2-1, x_2+2\nu_1+2\nu_2-1; x_1),
\end{gather*}
with
\begin{gather*}
\omega_{x_1}\sigma_{n_1}=\langle \nu_{23}(0), \nu_{123}, \nu|\nu_{12}(x_1), \nu_{123}, \nu\rangle\langle \nu_{23}(n_1), \nu_{123}, \nu|\nu_{12}(0), \nu_{123},  \nu\rangle,
\end{gather*}
where the following shorthand is used to show the dependence on the integers $x_1$ and $n_1$:
\begin{gather*} \nu_{23}(n_1)=n_1+\nu_2+\nu_3, \qquad \nu_{12}(x_1)=x_1+\nu_1+\nu_2.\end{gather*}
The operators are then
\begin{gather*}Q^{(12)}=\kappa(x_1, \beta_1),\qquad
 Q^{(23)}=(\omega_{x_1}\sigma_{n_1})^{1/2}\Lambda(x_1, \beta, x_2)(\omega_{x_1}\sigma_{n_1})^{-1/2}\equiv\hat{ \Lambda}(x_1, \beta, x_2).
 \end{gather*}

\subsection[The basis for $\big\{ Q^{(23)}, Q^{(234)}\big\}$ and bivariate Racah polynomials]{The basis for $\boldsymbol{\big\{ Q^{(23)}, Q^{(234)}\big\}}$ and bivariate Racah polynomials}

Since we know already \cite{KMPost11} that the representation of the algebra generated by the operators is associated with the bivariate Racah polynomials of Tratnik, we use this model to build a~representation of the algebra. In particular, we will identify the operators~$Q^{(ij)}$ and their linear combinations in terms of the eigenvalue and recurrence operators found for the Racah polynomials by Geronimo and Iliev~\cite{GI2010}.

 The (normalized) bivariate Racah polynomials are def\/ined to be
\begin{gather}
R_2(n_1,n_2; x_1,x_2; \beta;N) = \frac{r_{n_1}(\beta_1-\beta_0-1, \beta_2-\beta_1-1, -x_2-1,\beta_1+x_2; x_1)}{(-N)_{n+m}(-N+\beta_0)_{n+m}(\beta_2-\beta_1)_{n}(\beta_3-\beta_1)_{m}}
\nonumber\\
 \qquad{}\times r_{n_2}(2n_1+\beta_2-\beta_0-1, \beta_3-\beta_2-1, n_1-N-1, N+n_1+\beta_2; x_2-n_1).\label{2vracah}
\end{gather}
Note that we have dropped the hat notation from \cite{GI2010} since we have no need for the unnormalized Racah polynomials.
The polynomials (\ref{2vracah}) satisfy the following eigenvalue equations
\begin{gather*}
 \mathcal{L}_1^x R_2({\bf n; x}; \beta;N)=-n_1(n_1+\beta_2-\beta_0-1) R_2({\bf n; x}; \beta;N),\\
  \mathcal{L}_2^x R_2({\bf n; x}; \beta;N)=-(n_1+n_2)(n_1+n_2+\beta_2-\beta_0-1) R_2({\bf n; x}; \beta;N),
\end{gather*}
 with
\begin{gather*} \mathcal{L}_1^x=C_{(1,0)}^{(1)}(T_{x_1}-1)+C_{(-1,0)}^{(1)}\big(T_{x_1}^{-1}-1\big),\qquad
  \mathcal{L}_2^x=\sum_{j,k=0, \pm 1} C_{(j,k)}^{(2)}\big(T_{x_1}^jT_{x_2}^k-1\big),\\
C_1^{(1)}=\frac{(x_1+\beta_1-\beta_0)(x_1+\beta_1)(x_2+x_1+\beta_2)(x_2-x_1)}{(2x_1+\beta_1)(2x_1+\beta_1+1)},\\
C_{(1,1)}^{(2)}=\frac{(x_1+\beta_1)(x_1+\beta_1-\beta_0)(x_2+x_1+\beta_2)(x_2+x_1+\beta_2+1)(N-x_2)
(N+x_2+\beta_3)}{(2x_1+\beta_1)(2x_1+\beta_1+1)(2x_2+\beta_2)(2x_2+\beta_2+1)},\\
C_{(1,0)}^{(2)}=\frac{(x_1+\beta_1)(x_1+\beta_1-\beta_0)(x_2-x_1)(x_2+x_2+\beta_2)}
{(2x_1+\beta_1)(2x_1+\beta_1+1)}\\
\hphantom{C_{(1,0)}^{(2)}=}{}\times \frac{(2x_2(x_2+\beta_2)+2N(N+\beta_3)+(\beta_2+1)(\beta_3-1))}
{(2x_2+\beta_2-1)(2x_2+\beta_2+1)},\\
C_{(0,1)}^{(2)}=\frac{(2x_1(x_1+\beta_1)+(\beta_0+1)(\beta_1-1))(x_2+x_1+\beta_2)}
{(2x_1+\beta_1-1)(2x_1+\beta_1+1)}\\
\hphantom{C_{(0,1)}^{(2)}=}{}\times
\frac{(x_2-x_1+\beta_2-\beta_1)(N-x_2)(N+x_2+\beta_3)}
{(2x_2+\beta_2)(2x_2+\beta_2+1)},
\end{gather*}
and the other coef\/f\/icients def\/ined via the inversion operators
$I_1(f(x_1, x_2))=f(-x_1-\beta_1, x_2)$ and $I_2(f(x_1, x_2))=f(x_1, -x_2-\beta_2)$ as
\begin{gather*}
I_1\big(C^{(\ell)}_{(j,k)}\big)=C^{(\ell)}_{(-j,k)}, \qquad I_2\big(C^{(\ell)}_{(j,k)}\big)=C^{(\ell)}_{(j,-k)}.
\end{gather*}
For these operators and more, see~\cite[Appendix~A]{GI2010}.

Suppose that the expansion coef\/f\/icients can be expressed as
\begin{gather*}
P({\bf n}; {\bf x}; {\bf \beta};N)\equiv  \langle \nu_{23}, \nu_{234}, \nu|\nu_{12}, \nu_{123}, \nu\rangle=\sqrt{\omega_{{\bf x}} \sigma_{{\bf n}}}R_2({\bf n}; {\bf x}; {\bf \beta};N).
\end{gather*}
In terms of the quantum mechanical system (\ref{H}), these coef\/f\/icients correspond to the expansion coef\/f\/icients between eigenfunctions separated in spherical coordinates associated with diagonalizing the operators $\{Q^{(12)}, Q^{(123)}\}$ and the Hamiltonian, related to $Q$, into one where $\{Q^{(23)}, Q^{(234)}\}$ are diagonal.
The actions of $Q^{(12)}$, $Q^{(23)}$, $Q^{(123)}$ on both of the bases $|\nu_{12}, \nu_{123}, \nu\rangle $ and $|\nu_{23}, \nu_{234}, \nu\rangle$  can be determined from the results of the previous section. Using their actions, we def\/ine the operators $K_{1}$, $K_2,$ and $K_3$ that act on the expansion coef\/f\/icients as
\begin{gather*} K_1 P({\bf n}; {\bf x}; {\bf \beta};N) \equiv -\frac12 \big\langle \nu_{23}, \nu_{234}, \nu|Q^{(12)}|\nu_{12}, \nu_{123}, \nu\big\rangle
 =  -\frac12 \kappa\left( x_1, \frac{\beta_1+1}{2}\right) P({\bf n}; {\bf x}; {\bf \beta};N),\\
K_2 P({\bf n}; {\bf x}; {\bf \beta};N) \equiv -\frac12 \big\langle \nu_{23}, \nu_{234}, \nu|Q^{(23)}|\nu_{12}, \nu_{123}, \nu\big\rangle
 =  -\frac12 \hat{\Lambda}_1^{{\bf x}}({\bf x}, \beta, N) P({\bf n}; {\bf x}; {\bf \beta};N)\\
\hphantom{K_2 P({\bf n}; {\bf x}; {\bf \beta};N) }{}  =  -\frac12 \kappa\left( n_1, \frac{\beta_2-\beta_0}{2}\right)P({\bf n}; {\bf x}; {\bf \beta};N),\\
 K_3 P({\bf n}; {\bf x}; {\bf \beta};N) \equiv -\frac12 \big\langle \nu_{23}, \nu_{234}, \nu|Q^{(123)}|\nu_{12}, \nu_{123}, \nu\big\rangle
 =  -\frac12 \kappa\left( x_2, \frac{\beta_2+1}{2}\right) P({\bf n}; {\bf x}; {\bf \beta};N).
\end{gather*}
The shift operator in $x_1$ is the same as above (\ref{Lambda}) with $N$ replaced by $x_2$ and is given by
\begin{gather*} \Lambda_1^{{\bf x}}({\bf x}, \beta, N)\equiv \left[-\mathcal{L}_1^{{\bf x}}({\bf x}, \beta; N) +\left(\frac{\beta_2-\beta_0}{2}\right)\left(\frac{\beta_2-\beta_0}{2}-1\right)\right],
\end{gather*}
with the hat indicating conjugation by square root of the ground state $\sqrt{\omega_{{\bf x}} \sigma_{{\bf n}}}.$

The action of the operator $Q^{(234)}$ on the expansion coef\/f\/icients is similarly determined  leading to the def\/inition of a fourth operator $K_4$ satisfying
\begin{gather*}
K_4   P({\bf n}; {\bf x}; {\bf \beta};N) \equiv -\frac12 \big\langle \nu_{23}, \nu_{234}, \nu|Q^{(234)}|\nu_{12}, \nu_{123}, \nu\big\rangle \\
\hphantom{K_4   P({\bf n}; {\bf x}; {\bf \beta};N)}{}
 =  -\frac12 \kappa\left(n_1+n_2, \nu_2+\nu_3+\nu_4\right) P({\bf n}; {\bf x}; {\bf \beta};N) .
 \end{gather*}
However, the action of $Q^{(234)}$, as well as $Q^{(34)},$ on the basis $|\nu_{12}, \nu_{123}, \nu\rangle,$ or equivalently as shift operators in the variable $x_1$ and $x_2,$ have not yet been determined. Matching the eigenvalues of~$K_4$ leads to the identif\/ication
\begin{gather*}
K_4 P({\bf n}; {\bf x}; {\bf \beta};N)\equiv   -\frac12 \hat{\Lambda}_2^{{\bf x}}({\bf x}, \beta, N) P({\bf n}; {\bf x}; {\bf \beta};N),
\end{gather*}
where
\begin{gather*} \Lambda_2^{{\bf x}}({\bf x}, \beta, N) \equiv \left[ -\mathcal{L}_2^{{\bf x}}({\bf x}, \beta; N) +\left(\frac{\beta_3-\beta_0}{2}\right)\left(\frac{\beta_3-\beta_0}{2}-1\right)\right],
\end{gather*}
with the remaining $\beta$ given by
\begin{gather*}
\beta_3= 2\nu_1+2\nu_2+2\nu_3+2\nu_4-1.
\end{gather*}
The f\/inal operator to be determined is~$K_5, $ associated with the action of~$Q^{(34)}$. Since~$Q^{(34)}$ commutes with~$Q^{(12)}$, its action on the basis will be to shift the variable~$x_2$. We hypothesize that it is of the form
\begin{gather*} K_5   P({\bf n}; {\bf x}; {\bf \beta};N)\equiv -\frac12 \big\langle \nu_{23}, \nu_{234}, \nu|Q^{(34)}|\nu_{12}, \nu_{123}, \nu\big\rangle,
\end{gather*}
with
\begin{gather*} K_5   P({\bf n}; {\bf x}; {\bf \beta};N)\equiv   -\frac12 (\omega_{{\bf x}} \sigma_{{\bf n}})^{1/2}\Omega_1({\bf x}, \beta, N)(\omega_{{\bf x}} \sigma_{{\bf n}})^{-1/2} P({\bf n}; {\bf x}; {\bf \beta};N),\\
  \Omega_1({\bf x}, \beta, N)= -\tilde{B}({\bf x}) T_{x_2}-\tilde{E}({\bf x})T_{x_2}^{-1}+\tilde{B}({\bf x})+\tilde{E}({\bf x})+\frac{(\nu_3+\nu_4)(\nu_3+\nu_4-1)}{4}.
   \end{gather*}
The exact form of $\tilde{B}({\bf x})$ and $\tilde{E}({\bf x})$ are determined from the quadratic algebra relations, and are given by
\begin{gather*}
\tilde{B}({\bf x})=\frac{(x_2+x_1+\beta_2)(x_2-x_1+\beta_2-\beta_1)(N-x_2)(x_2+N+\beta_3)}{(2x_2+\beta_2+1)(2x_2+\beta_2)},\\
\tilde{E}({\bf x})=\frac{(x_2-x_1)(x_2+x_1+\beta_2)(N-x_2+\beta_3-\beta_2)(N+x_2+\beta_2)}{(2x_2+\beta_2-1)(2x_2+\beta_2)}.
\end{gather*}
Thus, we have realized each of the six linearly independent quadratic Casimirs $Q^{(ij)}$ in terms of the following operators, which act on the basis $|\nu_{12}, \nu_{123}, \nu \rangle$,
\begin{gather}
\label{Q12f} Q^{(12)} =   \kappa(x_1, \beta_1),\\
     Q^{(23)} = (\omega_{{\bf x}} \sigma_{{\bf n}})^{1/2}\Lambda_1^{{\bf x}}({\bf x}, \beta, N)(\omega_{{\bf x}} \sigma_{{\bf n}})^{-1/2},\\
Q^{(123)} =  \kappa(x_2, \beta_2),\\
    Q^{(234)} = (\omega_{{\bf x}} \sigma_{{\bf n}})^{1/2}\Lambda^{{\bf x}}_2({\bf x}, \beta, N)(\omega_{{\bf x}} \sigma_{{\bf n}})^{-1/2},\\
Q^{(1234)} = \kappa(N, \beta_3),\\
     Q^{(34)} = (\omega_{{\bf x}} \sigma_{{\bf n}})^{1/2}\Omega^{{\bf x}}_1({\bf x}, \beta, N)(\omega_{{\bf x}} \sigma_{{\bf n}})^{-1/2}.\label{Q34f}
     \end{gather}
Note that the algebra relations do not close without the addition of the shift operator representing~$Q^{(34)}$.

\section{Some consequences of the model}\label{section4}
In this section, we derive several results concerning the bivariate Racah polynomials which arise as a result of the above theory.

\subsection[The algebra ${\rm QR}(9)$]{The algebra $\boldsymbol{{\rm QR}(9)}$}

By construction, the operators (\ref{Q12f})--(\ref{Q34f}) give simply another basis for the quadratic Casimir operators and so the algebra generated by these operators is identical to the algebra def\/ined by~(\ref{quadsameijQijk}),~(\ref{quadsameij}). As we shall see in this section, this algebra is an extension of the algebra~${\rm QR}(3)$ with 9~linearly independent generators associated with the bivariate Racah polynomials and so we call it~${\rm QR}(9)$. In order to interpret the algebra as an extension of~${\rm QR}(3)$, we chose the following basis
\begin{alignat}{3}
\label{K1} &K_1=-\frac12\kappa(x_1, \beta_1), \qquad && K_2=-\frac12 \Lambda^{{\bf x}}_1({\bf x}, \beta, y),& \\
&K_3=-\frac12\kappa(x_2, \beta_2),\qquad && K_4=-\frac12 \Lambda^{{\bf x}}_2({\bf x}, \beta, N),& \\
&K_5=-\frac12\Omega_1^{{\bf x}}({\bf x}, {\beta}, N). \qquad && & \label{K5}
\end{alignat}
Here we have dropped the normalization and weight factors since they have no bearing on the algebra relations. Note that whereas the algebra generate by the operators~$Q^{(ij)}$ had 6 linearly independent generators, we have restricted to components with a f\/ixed~$N$ and so~$Q$, representing the total Casimir operator, will be considered as a~constant of the algebra and hence gives a~linear dependence between the 6 generators~(\ref{totalQ}). This is analogous to the univariate case and the algebra~${\rm QR}(3)$.

As we shall see, there will be several copies of the algebra ${\rm QR}(3)$ in this algebra. The f\/irst being the one generated by~$K_1$ and~$K_2$ since these are simply the operators which act on the f\/irst factor of $R_2({\bf n}; {\bf x}; \beta, N)$. In order to obtain this algebra, we include the commutator of~$K_1$ and~$K_2$ calling it~$L_1$. It would be $K_3$ in ${\rm QR}(3)$. The copy of the algebra~${\rm QR}(3)$ is then given~by
\begin{gather*}
 [ K_1, K_2 ] = L_1,\\
 [K_2, L_1 ] = K_2^2+\{K_1, K_2\} +d_1 K_2 +e_{11},\qquad
 [L_1, K_1 ] = K_1^2+\{K_1, K_2\} +d_1 K_1 +e_{12},
\\
 d_1=\frac12\big(Q^{(1)}+Q^{(2)}+Q^{(3)}-2K_3\big), \\
e_{11}=-\frac{1}4\big(Q^{(3)}-Q^{(2)}\big)\big(2K_3+Q^{(1)}\big), \qquad
e_{12}=\frac{1}4\big(Q^{(2)}-Q^{(1)}\big)\big(2K_3+Q^{(3)}\big) .
\end{gather*}
Notice that both~$K_1$ and~$K_2$ commute with~$K_3$ so~$d_1$, $e_{11}$, and $e_{12}$ act as constant for the sub-algebra generated by~$K_1$,~$K_2$ and~$L_2$.

We continue taking commutators and see
\begin{gather*}
[K_1, K_3]=[K_2, K_3]=0
\end{gather*}
but $[K_1, K_4]\ne 0$ and so we call it~$L_2$. The algebra generated by these operators closes again to form a copy of~${\rm QR}(3)$ but only if the operator $K_5$ is included in the list of generators. The algebra relations become
\begin{gather*}
 [K_1, K_3 ] = L_2,\\
 [K_4, L_2 ] = K_4^2+\{K_1, K_4\}+d_2K_4+e_{21},\qquad
 [L_2, K_1 ]  = K_1^2+\{K_1, K_4\}+d_2K_1+e_{22},\\
d_2=\frac{1}{2}\big(2K_5-Q^{(1)}-Q^{(2)}\big), \\
e_{21}=-\frac{1}{4}\big(Q-Q^{(1)}\big)\big(2K_5+Q^{(2)}\big), \qquad
e_{22}=\frac{1}{4}\big(Q^{(1)}-Q^{(2)}\big)(2K_5+Q) .
\end{gather*}

The operator $K_5$ is a shift operator in the variable~$x_2$ only so we expect it to generate, with~$K_3$ another copy of~${\rm QR}(3).$ We shall see in the next section that this is due to the fact that it is possible to diagonalize~$K_5$ with univariate Racah polynomials in the variable~$x_2$. Def\/ining
\begin{gather*}
[K_3, K_5]=L_3,
\end{gather*}
gives
\begin{gather*}
 [K_5, L_3 ] = K_5^2+\{K_3, K_5\} +d_3 K_3 +e_{31},\qquad
 [L_3, K_3 ] = K_3^2+\{K_3, K_5\} +d_3 K_4 +e_{32},
\\
 d_{3}=\frac12\big(Q-2K_1+Q^{(3)}+Q^{(4)}\big), \\
e_{31}=-\frac{1}{4}\big(Q^{(3)}-Q^{(4)}\big)(Q+2K_1), \qquad
e_{32}=\frac{1}{4}\big(Q-Q^{(4)}\big)\big(2K_1+Q^{(3)}\big).
\end{gather*}
Again, note that~$K_1$ commutes with both~$K_3$ and~$K_5$.

The f\/inal copy of ${\rm QR}(3)$ is generated by $K_2$ and $K_5$ and it is
\begin{gather*} L_4\equiv [K_2, K_5],\\
[K_5, L_4]=K_5^2+\{K_2, K_5\}+d_4K_4+e_{41},\qquad
 [L_4, K_2 ] = K_2^2+\{K_2, K_4\}+d_4K_2+e_{42},\\
d_4=-\frac{1}{2}\big(Q^{(2)}+Q^{(3)}+Q^{(4)}-2K_4\big), \\
e_{41}=-\frac{1}{4}\big(Q^{(3)}-Q^{(4)}\big)\big(Q^{(2)}+2K_4\big), \qquad
e_{42}=\frac{1}{4}\big(Q^{(2)}-Q^{(3)}\big)\big(Q^{(4)}+2K_4\big).
\end{gather*}
As in the previous cases, $K_4$ commutes with both $K_2$ and $K_5$ so acts as a constant in this subalgebra.

The operators $L_1$, $L_2$, $L_3$ and $L_4$ exhaust the set of linearly independent commutators of the~$K_j$'s. The only remaining non-zero commutator is
\begin{gather*}
[K_3, K_4]=L_4+L_3-L_2-L_1.
\end{gather*}
The remaining algebra relations def\/ining ${\rm QR}(9)$ can, as with the previous ones, be directly determined by~(\ref{quadsameijQijk}),~(\ref{quadsameij}) but we give a set for example. They are
\begin{gather*}
 [K_3, L_1 ] = 0,\\
  [K_4, L_1 ] = \frac12\big(\{K_1, K_2\}+\{K_1, K_4\}+\{K_2, K_4\}+\{K_3, K_5\}-\{K_2, K_5\}-\{K_3, K_5\}\big)\\
 \hphantom{[K_4, L_1 ] =}{}
+\frac{Q^{(4)}}{2}K_1+\frac{Q^{(1)}}{2}(K_2+K_5)+\frac{Q^{(2)}}{2}(K_3+K_4)\\
\hphantom{[K_4, L_1 ] =}{}
+\frac{1}{4}\big(Q^{(1)}Q^{(2)}+Q^{(1)}Q^{(4)}+Q^{(2)}Q^{(4)}\big),\\
 [K_5, L_1 ] = \frac12\big(\{K_1, K_4\}+\{K_3, K_5\}-\{K_1, K_2\}-\{K_1, K_5\}-\{K_3, K_4\}-\{K_2, K_5\}\big)\\
\hphantom{[K_5, L_1 ] = }{}
-\frac{Q^{(3)}}2(K_1+K_4)-\frac{Q^{(2)}}{2}(K_3+K_5)-\frac{Q}{2}K_2+\frac14\big(Q^{(2)}+Q^{(3)}\big)Q+\frac14Q^{(2)}Q^{(3)}.
\end{gather*}
To summarize, as in the univariate case, the algebra generated by the eigenvalue operators and multiplication by the variables of the bivariate Racah polynomials can be closed to form a~quadratic algebra. But unlike the univariate case, an additional operator not arising as a~commutator of two basis operators  must be adjoined to the algebra. This operator is a shift operator in the second variable~$x_2.$  With this additional operator, the algebra closes and has 9~generators, the operators $K_1$ through $K_5$ and their 4~linearly independent commutators. The algebra contains at least 4 subalgebras isomorphic to ${\rm QR}(3)$. Finally, we emphasize that this algebra is isomorphic to the algebra generated by the operators~$Q^{(ij)}$ whose algebra is much more compactly expressed via~(\ref{quadsameijQijk}),~(\ref{quadsameij}). The form of the algebra  in terms of the~$K_i$ is given here to emphasize that it is a natural extension of the algebra~${\rm QR}(3)$.

\subsection{Duality in the model}
As another consequence of the model, we describe the duality of the operators mentioned above in terms of the expansion coef\/f\/icients. Recall, that we have
\begin{gather*} \big\langle \nu_{23}, \nu_{234}, \nu | Q^{(12)}|\nu_{12}, \nu_{123},\nu \big\rangle = \kappa\left(x_1, \frac{{\beta}_1+1}{2}\right) \langle \nu_{23}, \nu_{234}, \nu |\nu_{12}, \nu_{123},\nu \rangle,\\
\big\langle \nu_{23}, \nu_{234}, \nu | Q^{(123)}|\nu_{12}, \nu_{123},\nu \big\rangle = \kappa\left(x_2, \frac{{\beta}_2+1}{2}\right)\langle \nu_{23}, \nu_{234}, \nu |\nu_{12}, \nu_{123},\nu \rangle,\\
\big\langle \nu_{23}, \nu_{234}, \nu | Q^{(23)}|\nu_{12}, \nu_{123},\nu \big\rangle = \hat{\Lambda}_1^x\langle \nu_{23}, \nu_{234}, \nu |\nu_{12}, \nu_{123},\nu \rangle,\\
\big\langle \nu_{23}, \nu_{234}, \nu | Q^{(234)}|\nu_{12}, \nu_{123},\nu \big\rangle =
\hat{\Lambda}_2^x\langle \nu_{23}, \nu_{234}, \nu |\nu_{12}, \nu_{123},\nu \rangle.
\end{gather*}
The operator $Q^{(12)}$ can of course act on the basis $|\nu_{23}, \nu_{234}\rangle$ and will give a shift operator of $0, \pm 1$ on each of the values~$\nu_{23}$ and~$\nu_{234}.$ Similarly, the operator~$Q^{(123)}$ will act on the basis $|\nu_{23}, \nu_{234}\rangle$ however it will only shift the value~$\nu_{234}$. We can write these as shift operators in~$n_1$ and~\mbox{$n_1+n_2$}~as
\begin{gather*}
\big\langle \nu_{23}, \nu_{234}, \nu | Q^{(12)}|\nu_{12}, \nu_{123},\nu \big\rangle = \hat{\Lambda}_2^n \langle \nu_{23}, \nu_{234}, \nu |\nu_{12}, \nu_{123},\nu \rangle,\\
\big\langle \nu_{23}, \nu_{234}, \nu | Q^{(123)}|\nu_{12}, \nu_{123},\nu \big\rangle = \hat{\Lambda}_1^n\langle \nu_{23}, \nu_{234}, \nu |\nu_{12}, \nu_{123}.\nu \rangle.
\end{gather*}

These two operators hence form a set of commuting shift operators for the same expansion coef\/f\/icients but with a new set of variables $\tilde{x}$, degrees $\tilde{n}$ and coef\/f\/icients~$\tilde{\beta}. $ In the new variables the expansion coef\/f\/icients are
\begin{gather*}
\langle \nu_{23}(\tilde{x}_2), \nu_{234}(\tilde{x}_1), \nu |\nu_{12}(\tilde{n}_1+\tilde{n}_2), \nu_{123}(\tilde{n}_1),\nu \rangle=\sqrt{\omega_{{\bf \tilde x}}\sigma_{\bf \tilde{n}}}{R}_2 ({\bf \tilde{n}; \tilde{x}; \tilde{\beta}; }N ),
\end{gather*}
with action of the operators as
\begin{gather*}
\big\langle \nu_{23}, \nu_{234}, \nu | Q^{(12)}|\nu_{12}, \nu_{123},\nu \big\rangle = \hat{\tilde{\Lambda}}_2^x\langle \nu_{23}, \nu_{234}, \nu |\nu_{12}, \nu_{123},\nu \rangle,\\
\big\langle \nu_{23}, \nu_{234}, \nu | Q^{(123)}|\nu_{12}, \nu_{123},\nu \big\rangle = \hat{\tilde{\Lambda}}_1^x\langle \nu_{23}, \nu_{234}, \nu |\nu_{12}, \nu_{123},\nu \rangle,\\
\big\langle \nu_{23}, \nu_{234}, \nu | Q^{(23)}|\nu_{12}, \nu_{123},\nu \big\rangle = \kappa\left(\tilde{x}_2, \frac{\tilde{\beta}_2+1}{2}\right)\langle \nu_{23}, \nu_{234}, \nu |\nu_{12}, \nu_{123},\nu \rangle,\\
\big\langle \nu_{23}, \nu_{234}, \nu | Q^{(234)}|\nu_{12}, \nu_{123},\nu \big\rangle = \kappa\left(\tilde{x}_1, \frac{\tilde{\beta}_1+1}{2}\right)\langle \nu_{23}, \nu_{234}, \nu |\nu_{12}, \nu_{123},\nu \rangle.
\end{gather*}
The equations necessary to satisfy this duality are then
\begin{gather}\label{dual1}
\kappa\left(x_1, \frac{\beta_1+1}{2}\right)=\kappa\left(\tilde{n}_1+\tilde{n}_2, \frac{\tilde{\beta}_3-\tilde{\beta}_0}{2}\right), \qquad
 \kappa\left(x_2, \frac{\beta_2+1}{2}\right)=\kappa\left(\tilde{n}_1, \frac{\tilde{\beta}_2-\tilde{\beta}_0}{2}\right), \\
\kappa\left({n}_1+{n}_2, \frac{{\beta}_3-{\beta}_0}{2}\right)= \kappa\left(\tilde{x}_1, \frac{\tilde{\beta}_1+1}{2}\right), \qquad
\kappa\left({n}_1, \frac{{\beta}_2-{\beta}_0}{2}\right)= \kappa\left(\tilde{x}_2, \frac{\tilde{\beta}_2+1}{2}\right),\label{dual2}
\end{gather}
as well as
\begin{gather} \label{dualop1}
\Lambda_1^n=\hat{\Lambda}_1^x \Rightarrow \left[\kappa\left(N, \frac{\beta_2+1}{2}\right) -\mathcal{L}_1^n\right]= \left[\kappa\left(0, \frac{\tilde{\beta}_3-\tilde{\beta}_0}{2}\right) -\tilde{\mathcal{L}}_1^x\right],\\
\Lambda_2^n=\hat{\Lambda}_2^x \Rightarrow \left[\kappa\left(N, \frac{\beta_1+1}{2}\right) -\mathcal{L}_2^n\right]=\left[\kappa\left(0, \frac{\tilde{\beta}_2-\tilde{\beta}_0}{2}\right) -\tilde{\mathcal{L}}_2^x\right], \label{dualop2}
\end{gather}
where $\mathcal{L}_j^n$ are some operators of the form
\begin{gather*}  
\mathcal{L}_j^n =\sum_{j+k=0, \pm 1 } D_{(j,k)}^{(2)}\big(T_{n_1}^jT_{n_2}^k-1\big).
\end{gather*}
Solving~(\ref{dual1}),~(\ref{dual2}) as well as the constant terms of~(\ref{dualop1}),~(\ref{dualop2}) gives exactly the dual variables def\/ined in \cite{GI2010}, namely
\begin{alignat*}{3}
& x_1=\tilde{n}_1+\tilde{n}_2+\tilde{\beta}_3-\tilde{\beta}_0+N-1, \qquad && x_2= \tilde{n}_1+\tilde{\beta}_2-\tilde{\beta}_0+N-1,& \\
& n_1=\tilde{x}_2+\tilde{\beta}_2+N, \qquad && n_2=\tilde{x}_1-\tilde{x}_2+\tilde{\beta}_1-\tilde{\beta}_2, & \\
& \beta_0=\tilde{\beta}_0, \qquad && \beta_1=\tilde{\beta}_0-\tilde{\beta}_3-2N+1,& \\
& \beta_2=\tilde{\beta}_0-\tilde{\beta}_2-2N+1, \qquad && \beta_3=\tilde{\beta}_0-\tilde{\beta}_1-2N+1.&
\end{alignat*}
From these identif\/ications, we are able to identify the shift operators as $ \mathcal{L}_j^n=\tilde{\mathcal{L}}_j^x,$ in agreement with~\cite{GI2010}.
Thus, we see that the duality and therefore the bispectrality of the bivariate Racah polynomials arise from symmetries implicit in the model.

\subsection[The action of  $Q^{(34)}$ and another set of  bivariate Racah polynomials]{The action of  $\boldsymbol{Q^{(34)}}$ and another set of  bivariate Racah polynomials}

As mentioned above, the operators $K_3$ and $K_5$ generate a copy of the algebra~${\rm QR}(3)$ and so we anticipate that they will be associated with Racah polynomials as well. Ignoring initially the normalization factors, let us consider simply the set of commuting operators
$\{ \Omega_1^x({\bf x}, \beta, N)$, $\Lambda_2^x({\bf x}, \beta, N)\}$. The operator~$ \Omega_1^x({\bf x}, \beta, N)$ has the form of an eigenvalue operator for a univariate Racah polynomial in the variable $x_2-x_1$. However, in this new variable the operator $\Lambda_2^x({\bf x}, \beta, N)$ would include shifts in this variable of $x_2-x_1\rightarrow x_2-x_1+2$. With this observation in mind, we interpret $\Omega_1^x({\bf x}, \beta, N)$ as a shift operator in $\tilde{m}_2=x_2-x_1$. Indeed, making the identif\/ications
\begin{gather*} \tilde{m}_1=x_1, \qquad   \tilde{m}_2=x_2-x_1,\\
\tilde{\gamma}_0=-2N-2\nu_1-2\nu_2-2\nu_3-2\nu_4+1,\\
\tilde{\gamma}_1=-2N-2\nu_2-2\nu_3-2\nu_4+1,\\
\tilde{\gamma}_2=-2N-2\nu_3-2\nu_4+1,\\
\tilde{\gamma}_3=-2N-2\nu_4+1,
\end{gather*}
the two commuting operators can be realized as
\begin{gather*}
\Omega_1^{{\bf x}}({\bf x}, \beta, N) = \left(N+\frac{\tilde{\gamma}_2+1}{2}\right)\left(N+\frac{\tilde{\gamma}_2-1}{2}\right)-{\mathcal{L}}_1^{{\bf \tilde{m}}}
 \equiv   \Lambda_1^{\tilde{m}}({\bf \tilde{m}}, \gamma, N),\\
\Lambda_2^{{\bf x}}({\bf x}, \beta, N) = \left(N+\frac{\tilde{\gamma}_1+1}{2}\right)\left(N+\frac{\tilde{\gamma}_1-1}{2}\right)-{\mathcal{L}}_2^{{\bf \tilde{m}}}
 \equiv   \Lambda_2^{\tilde{m}}({\bf \tilde{m}}, \gamma, N).
 \end{gather*}
A set of simultaneous eigenvectors for these operators are the normalized, Racah polynomials ${R}_2({\bf \tilde{m}; \tilde{y};} \tilde{\gamma}; N)$ satisfying the eigenvalues equations
\begin{gather*} \Lambda_1^{\tilde{m}}{R}_2({\bf \tilde{m}; \tilde{y};} \tilde{\gamma}; N)=\kappa\left(\tilde{y}_2, \frac{\tilde{\gamma}_2+1}{2}\right){R}_2({\bf \tilde{m}; \tilde{y};} \tilde{\gamma}; N),\\
     \Lambda_2^{\tilde{m}}{R}_2({\bf \tilde{m}; \tilde{y};} \tilde{\gamma}; N)=\kappa\left(\tilde{y}_1, \frac{\tilde{\gamma}_1+1}{2}\right){R}_2({\bf \tilde{m}; \tilde{y};} \tilde{\gamma}; N).
     \end{gather*}
To return to $x_1$ and $x_2$ we transform to the dual variables, or rather as these were set up already as the dual variables, we transform back to the variables~$y_1$,~$y_2$ and~$m_1$,~$m_2.$ In particular, the polynomials can be represented as
\begin{gather*}
{R}_2({\bf \tilde{m}; \tilde{y}}; \tilde{\gamma}; N)={R}_2({\bf m; y}; {\gamma}; N),
\end{gather*}
with
\begin{alignat}{3}
& y_1=x_2+N+2\nu_1+2\nu_2+2\nu_3-1, \qquad && y_2=x_1+N+2\nu_1+2\nu_2-1,& \nonumber\\
&  m_1=\tilde{y}_2+\tilde{\gamma}_2+N, \qquad && m_2=\tilde{y}_1-\tilde{y}_2+\tilde{\gamma}_2-\tilde{\gamma}_2,& \label{ys}
\end{alignat}
   and
\begin{alignat*}{3}
& {\gamma}_0=-2N-2\nu_1-2\nu_2-2\nu_3-2\nu_4+1,\qquad && {\gamma}_1=-2N-2\nu_1-2\nu_2-2\nu_3+1,&\\
& {\gamma}_2=-2N-2\nu_1-2\nu_2+1,\qquad && {\gamma}_3=-2N-2\nu_1+1.&
\end{alignat*}

{\sloppy The eigenvalue equations for the polynomials in ${\bf y}$, def\/ined via (\ref{ys}), and ${\bf m}$ are
\begin{gather*}
\Omega_1^{{\bf x}}{R}_2({\bf m; {y}}; {\gamma}; N)=\kappa\left(m_1, \frac{{\gamma}_2-\gamma_0}{2}\right){R}_2({\bf m; y}; {\gamma}; N),\\
\Lambda_2^{{\bf x}}{R}_2({\bf m; y}; {\gamma}; N)=\kappa\left(m_1+m_2, \frac{{\gamma}_3-\gamma_0}{2}\right){R}_2({\bf m; y}; {\gamma}; N).
\end{gather*}
Thus, we see that the operator $\Omega_1^x({\bf x}, \beta, N)$ is another operator that commutes with $\Lambda_2^x({\bf x}, \beta, N)$ and the set of these operators are diagonalized by the bivariate Racah polynomials ${R}_2({\bf m, y}, {\gamma}, N)$.

}

To return to the model. The expansion coef\/f\/icients from the basis $|\nu_{12}, \nu_{123}, \nu\rangle $ to $|\nu_{34}, \nu_{234}, \nu\rangle$ can be expressed as
\begin{gather*}
\langle \nu_{34}(m_1), \nu_{234}(m_1+m_2), \nu| \nu_{12}(x_1), \nu_{123}(x_2), \nu\rangle =\sqrt{\omega_{\bf y} \sigma_{\bf m}} {R}_2(m_1, m_2; y_1,y_2; {\gamma}, N),
\end{gather*}
which are Racah polynomials in the variables (\ref{ys}). They can be represented as a product of one Racah polynomial of degree $m_1$ depending on both~$x_1$ and~$x_2$ and a~second one of degree~$m_2$ depending on~$x_1$ only.

\section{Conclusions}\label{section5}
In this article, we have seen how the bivariate Racah polynomials are related to dif\/ferent~$\mathfrak{su}(1,1)$ coupling schemes, in analogy with the univariate case~\cite{genest2013superintegrability}. The bivariate Racah polynomials are seen to be expansion coef\/f\/icients between sets of  eigenfunctions for the Hamiltonian (\ref{H}) in dif\/ferent choices of spherical coordinates. This Hamiltonian arises as the Casimir operator of  the tensor product of four copies of $\mathfrak{su}(1,1)$~\cite{genest2014generic} and its conserved quantities are seen to be represented in terms of intermediate Casimir operators for dif\/ferent coupling schemes.

As consequences of the model, we have shown that the algebra generated by multiplication by the arguments of the polynomials along with the eigenvalue operators for the bivariate Racah polynomials do not close to form a quadratic algebra, as in the univariate case, unless an additional operator is included, beyond the operators arising directly as a commutator of the generators. This additional operator is a shift operator in the second variable, $x_2$ and commutes with shift operator~$\Lambda_2^x$ (or equivalently $\mathcal{L}_2^x$). Including this additional operator, the algebra is generated by~5 operators plus their 4 linearly independent commutators. This algebra is isomorphic to the algebra generated by the intermediate Casimir operators for the tensor product of four $\mathfrak{su}(1,1)$ representations. However, taking the basis as in (\ref{K1})--(\ref{K5}) makes explicit that the algebra is an extension of the Racah algebra~${\rm QR}(3)$ and indeed contains at least four subalgebras isomorphic to ${\rm QR}(3).$ We have also seen how the symmetry of the Hamiltonian and the representation in terms of the tensor product allows for a natural def\/inition of the dual variables and bispectrality properties of the polynomials. Finally, from the model and the algebra relations we have identif\/ied an additional operator $\Omega_1^x$ that commutes with $\Lambda_2^x$ but is a shift operator in the second variable only. This additional operator is necessary for the quadratic algebra to close. The eigenfunctions of $\{\Omega_1^x, \Lambda_2^x\}$ are also bivariate Racah polynomials but of dif\/ferent arguments and with a dif\/ferent choice of parameters.

Let us f\/inish by mentioning several other possible consequences of this model that would be interesting to investigate. The f\/irst is to utilize more explicitly the separated eigenfunctions of the Hamiltonian (\ref{H}) to derive integral formulas and identities for the Racah polynomials, analogous to the results of~\cite{genest2014generic} for the~$9j$ coef\/f\/icients of the same model. A similar investigation has also been completed for the univariate case~\cite{miller2014wilson}. A second avenue is to extend the model into arbitrary dimensions. The $N$-dimensional analog of the quantum Hamiltonian~(\ref{H}) is clear and it remains superintegrable, exhibiting $N(N-1)/2$ linearly, though not functionally,  independent integrals of the motion via the~$Q^{(ij)}$'s~(\ref{Qijmodel}). Similarly, it is clear how to extend the algebraic interpretation in terms of the tensor product of $N$ copies of $\mathfrak{su}(1,1)$ representations. The algebra relations would be directly analogous to (\ref{quadsameijQijk}),~(\ref{quadsameij}), having been expressed in a~symmetric and almost dimensionally independent manner. The remaining open question would be whether the Tratnik polynomials~\cite{GI2010, Trat1991},  already def\/ined for arbitrary numbers of variables, are the appropriate $N-1$ variable extensions of the Wilson polynomials realizing this algebra. Just as in the case of the bivariate polynomials, we expect that additional operators beyond the~2~sets of $N-1$ commuting operators def\/ining the recurrence and eigenvalues relations will be necessary to for the algebra to close.

\looseness=-1
Another interesting avenue of research would be to consider the limits of these Racah polynomials in terms of the contractions of algebra. Again, in the univariate case, it has been shown that the limits of the Racah polynomials are generated by contractions of the quadratic algebras~\cite{KMPost13} and furthermore that all such contractions are in fact generated by Lie algebra contractions~\cite{kalnins2014quadratic}. For the bivariate case, some results have already been obtained. Equivalence classes of second-order superintegrable systems in conformally f\/lat space have been identif\/ied~\cite{CapelKress2014} as have the contractions between them~\cite{CapelKressPost2015}. It remains to identify the corresponding limits of the polynomials and the action on the algebras. As an example, the multivariate Hahn polynomials have been represented in a~similar manner~\cite{genest2014multivariate} corresponding to the singular harmonic oscillator in~3D, which can be obtained from the Hamiltonian~(\ref{H}) by an appropriate limit. The connection between these contraction limits, their quadratic algebras and dif\/ferent coupling schemes should provide a fruitful context for classifying multivariate orthogonal polynomials and their limits.

Finally, we mention that this analysis would also be interesting to extend to the $q$-case and to consider the appropriate algebra for the bivariate $q$-Racah polynomials~\cite{iliev2011bispectral} and the connections with Leonard triples~\cite{GWH}, the universal enveloping algebra $U_q(\mathfrak{sl}_2)$~\cite{terwilliger2011universal} and Hecke algebras~\cite{Koorn1, Koorn2}.

\subsection*{Acknowledgments}

This manuscript is in honor of Luc Vinet in celebration of his 60th birthday. The author would like to thank the referees for their invaluable comments and suggestions.

\pdfbookmark[1]{References}{ref}
\LastPageEnding

\end{document}